\documentclass[12pt,preprint]{aastex}
 
\slugcomment{To appear in \apj}
 
\shorttitle{The Role of Gas in the Merging of Massive Black Holes in Galactic Nuclei.}
\shortauthors{Escala, Larson, Coppi \& Mardones}

\def\msun{M_{\odot}}

\begin{document}
 
\title{The Role of Gas in the Merging of Massive Black Holes in Galactic Nuclei.  II. Black Hole Merging in a Clumpy Disk.} 
\author{Andr\'es Escala}
\affil{
Department of Astronomy, Yale University, New Haven, CT06520-8101, USA \& Departamento de Astronom\'{\i}a, Universidad de Chile, Casilla 36-D, Santiago, Chile.} 

\author{Richard B. Larson \& Paolo S. Coppi}
\affil{
Department of Astronomy, Yale University, New Haven, CT06520-8101, USA.}  

\author{Diego Mardones}
\affil{
Departamento de Astronom\'{\i}a, Universidad de Chile, Casilla 36-D, Santiago, Chile.}

\begin{abstract}   
Using high-resolution SPH numerical simulations, we investigate the effects of gas on the inspiral and merger of a massive black hole binary. This study is motivated by the  very massive nuclear gas disks observed in the central regions of merging galaxies. Here we  present results that expand on the treatment in a previous work (Escala, Larson, Coppi \& Mardones 2004; henceforth Paper I), by studying more realistic models in which the gas is in a disk with significant clumpiness. We run a variety of models, ranging from  simulations  with a  relatively smooth gas disk to cases in which the gas has a more clumpy spatial distribution. We also vary the inclination angle between the plane of the binary and the plane of the disk, and the  mass ratio between the MBHs and the gaseous disk. We find that as in Paper I, in the early evolution of the system the binary separation diminishes  due to gravitational drag, and in the later stages the medium responds by forming an ellipsoidal density enhancement whose axis lags behind the binary axis; this offset produces a torque on the binary that causes continuing loss of angular momentum and is able to reduce the separation to distances where gravitational radiation is efficient. The main difference is that between these two regimes we now find a new  transition regime that was not apparent in Paper I, in which the evolution is temporarily slowed down when neither of these mechanisms is fully effective. In the variety of simulations that we perform, we find that the coalescence timescale for the MBH binary varies between $\rm 5\times 10^{6} yr$ and $\rm 2.5\times 10^{7} yr$. For MBHs that satisfy the observed `$m-\sigma_{c}$' relation, we predict that in a merger of  galaxies that have at least 1\% of their total mass in gas, the MBHs will coalesce soon after the galaxies merge. We also predict that if the MBHs depart considerably  from the `$m-\sigma_{c}$' relation, strong tidal and/or resonant forces from the MBH binary can create a circumbinary gap in the disk that stalls the coalescence, but this gap formation can act as a self-regulatory mechanism on MBH growth that can help  to explain the existence of the `$m-\sigma_{c}$' relation. Our work thus supports scenarios of massive black hole evolution and growth in which hierarchical merging plays an important role. The final coalescence of the black holes leads to gravitational radiation emission that would be detectable out to high redshift by LISA. 
\end{abstract}  


\section{Introduction}  


The possibility of massive black hole (MBH) mergers follows from two widely accepted facts: that galaxies merge and that every galaxy with a significant bulge hosts an MBH at its center (Richstone et al. 1998). This problem was first considered by Begelman, Blandford, \& Rees (1980) in a study of the long-term evolution of a black hole binary at the center of a dense stellar system. Initially, dynamical friction brings the two black holes toward the center of the
system. The resulting binary MBH continues to shrink via 3-body interactions with the surrounding stars, but 3-body interactions tend to eject stars from the central region, causing the merger eventually to stall, unless some additional mechanism is able to extract angular momentum from the MBH binary.

The possible additional mechanism that extracts angular momentum from the binary is very likely to be gas dynamical in origin. Observations of gas-rich interacting galaxies such as the `Ultraluminous Infrared Galaxies' (ULIRGs) indicate that large amounts of gas are present in the central regions of merging galaxies (Sanders \& Mirabel 1996). This massive nuclear gas concentration has an important influence on the evolution of any central MBH binary that forms following a galaxy merger.  Since the gas is strongly dissipative, unlike the stars, it is expected to remain concentrated near the center and thus to play a continuing role in driving the evolution of a central binary MBH. Our aim is to study numerically the role of massive nuclear gas clouds  in driving the evolution of a binary MBH. 


In a previous paper (Escala, Larson, Coppi \& Mardones 2004; henceforth Paper I) we studied numerically the role of a massive spherical gas cloud in driving the evolution of a binary MBH. We followed the evolution of the binary through many orbits and close to the point where gravitational radiation becomes important. In Paper I, we presented results for a relatively simple idealized case in which the gas is assumed to be supported by a high virial temperature, so that the gas retains a nearly spherical and relatively smooth distribution. We found that in the early evolution of the binary, the separation decreases due to the gravitational drag exerted by the background gas. In the later stages, when the binary dominates the gravitational potential in its vicinity, the medium responds by forming an ellipsoidal density enhancement whose axis lags behind the binary axis, and this offset produces a torque on the binary that causes continuing loss of angular momentum and is able to reduce the binary separation to distances where gravitational radiation is efficient. Assuming typical parameters from observations of Ultra Luminous Infrared Galaxies, we predicted that a black hole binary will merge within $10^{7}$yrs.

The model presented in Paper I is relatively idealized and, as we describe in \S 2, observations suggest  that the gas in merging systems is in a rotating nuclear disk. In this paper, we study numerically the role of a massive gas disk, like those seen in the central regions of ULIRGs, in driving the evolution of a binary MBH. We run a variety of models, ranging from  simulations  with a  relatively smooth gas disk to cases in which the gas has a more clumpy spatial distribution. We follow the evolution of the binary through many orbits and close to the point where gravitational radiation becomes important.

We start with a review of the properties of the inner regions of merging galaxies in \S 2. We continue with a description of the assumed initial conditions and the model setup in \S 3. In \S 4 we study the evolution of a binary MBH in a massive gas disk, exploring the parameters that may be relevant for the coalescence of the  binary MBH. In \S 5 we study the final coalescence of the binary MBH using higher resolution simulations. In \S 6 we study the criteria for opening a circumbinary gap and the implications for the `$m-\sigma_{c}$' relation. Finally, our conclusions are presented in \S 7.

\section{Gas in the Nuclei of Merging Galaxies} 

Observational  and theoretical work both indicate that large amounts of gas can be present in the central regions of interacting galaxies, and that this gas can be a dominant component of these regions.  Numerical simulations show that in a merger of galaxies containing gas, much of the gas can be driven to the
center by gravitational torques that remove angular momentum from the
shocked gas (Barnes \& Hernquist 1992, 1996; Mihos \& Hernquist 1996); as a
result, more than 60\% of the gas originally present in the merging galaxies
can end up in a very massive nuclear disk with a radius of
several hundred parsecs (Barnes 2002).  
Observations of gas-rich interacting galaxies such
as the `Ultraluminous Infrared Galaxies' (ULIRGs) confirm that their central
regions often contain massive and dense clouds of molecular and atomic gas
whose masses are comparable to the total gas content of a large gas-rich
galaxy (Sanders \& Mirabel 1996).

Downes \& Solomon (1998, hereafter DS), using CO interferometer data, show that the molecular gas in ULIRGs, at subarcsecond resolution, is in  rotating nuclear disks. These molecular disks are highly turbulent, with typical  velocity dispersions of $\rm \sigma=100 \, km \, s^{-1}$. DS modeled the CO luminosity as coming mostly from a relatively low-density phase with a gas kinetic temperature in the range $\rm \sim 65-100 K$. This `low-density' gas is in a smooth medium and not in self-gravitating clouds, and has  an average density of $\rm 10^{3}cm^{-3}$. DS estimate that only $\sim$ 10\% of the gas is in highly opaque and dense ($\rm 10^{5}cm^{-3}$) self-gravitating clouds that are stable against tidal forces, although the exact fraction is still unclear.  The derived total gas mass is high, typically $\rm 5 \times 10^{9} \msun$, which is similar to the mass of molecular clouds in a large, gas-rich spiral galaxy. However, the total mass in molecular gas is still small compared to the dynamical mass derived from the rotation curve, the ratio of gas mass to enclosed dynamical mass being about $ \rm M_{gas}/M_{dyn}$ = 1/6.  In addition to these multi-phase nuclear disks, DS also found exceptionally massive clumps with ongoing extreme starburst activity having  characteristic sizes of only 100pc, with about $\rm 10^{9} \msun$ of gas and an IR luminosity of $\rm 3 \times 10^{11} L_{\sun}$ from recently formed OB stars.

Since ULIGRs are characterized by an ongoing starburst, feedback effects from star formation may play a major role in determining the physics of the ISM. The main feedback effects  on the gas  arise from massive stars and include stellar winds, supernova explosions, and the effect of radiation pressure. Recently, Wada and collaborators have made some attempts to model this complex environment including all the relevant feedback effects (Wada \& Norman 2001, Wada 2001). They found a multi-phase ISM in which substructure is continuously formed and destroyed, but it reaches a quasi-steady state which is characterized by a network of cold (T $<$ 100 K) dense clumps and filaments embedded in a hot (T $>\rm 10^{6}$ K) diffuse medium.

\section{Model Setup}

Having reviewed the observations of molecular gas in ULIGRs, we now describe our approach to modeling the complex gaseous environment in the inner regions of merging galaxies. 

In our model, the gas is distributed in a rotating disk with a Mestel surface density profile (Mestel 1963)
\begin{equation}  
\Sigma_{gas} (R) = \frac{\Sigma_{0} R_{0}}{R}    
\end{equation}
and it is initially distributed  uniformly  between z=$\pm \, 0.5$. In our simulations, we use the following units : [Mass] = $5 \times 10^{9}\msun$, [Time] =  2.5 $\times 10^{5}$yr, [Velocity] = 156.46 Km $s^{-1}$ and  [Distance] =  40 pc. In these units, the gravitational constant G is 22.0511, the disk radius is 10, and the total gas mass is 1.

In addition to the gaseous disk, in this paper we include a stellar bulge that stabilizes the disk and plays an important role in the evolution of the MBH binary when the binary is out of the plane of the disk. The bulge is modeled using a Plummer Law (Plummer 1911)
\begin{equation}
\rho (r) = \frac{3}{4} \frac{\rm M_{bulge}}{a^{3}} \left(1 +\frac{r^{2}}{a^{2}} \right)^{-5/2}   
\end{equation}
where $a$=200 pc is the core radius and $\rm  M_{bulge}$ is the total mass of the bulge. The  mass of the bulge within r=10  is 5 in the just mentioned units. This choice was made  to satisfy the observed rotation curves (Downes \& Solomon 1998) which imply that  the total gaseous mass is 1/6 of the dynamical mass. We model the bulge with a collection of 100,000 collisionless particles, with  a gravitational softening length of 4 pc. For the gaseous disk, we employ 235,331 SPH particles, and the softening length is also 4pc. 

We first considered a cold  massive disk (T=200K, $M_{disk}=5 \times 10^{9} \msun$) with an isothermal equation of state. This temperature is much smaller than the virial temperature, and therefore the cold disk fragments into clumps. These clumps  undergo a runaway collapse because of the isothermal equation of state, and the collapse  stops only at the gravitational softening length. The result is  the formation of  clumps with average densities many orders of magnitude higher than the observed ones (Downes \& Solomon 1998), and therefore the simulated clumps are physically unrealistic. In this unrealistic model, the MBHs interact violently and chaotically with the clumps, but in most cases quickly end up at the center where they continue to lose angular momentum by interaction with a central massive clump, qualitatively as in Paper I. 

In reality, in  the inner regions of ULIRGs, feedback processes from starburst activity may tend to halt the runaway collapse of such massive clumps. These  feedback effects and the detailed evolution of the clumps are very hard to model numerically in detail, so instead we adopt a more phenomenological approach, using a parametrized equation of state that allows us to represent a clumpy medium with the observed  range of densities. The detailed evolution of the clumps is not relevant for our problem because the gravitational drag effect depends strongly  on only one  property of the gas, its local density. Our motivation here is not to develop a fully realistic description of the gas, but only to properly model the orbital decay of the MBH binary caused by gravitational drag effects in a non-uniform medium.

 In order to prevent the formation of runaway clumps, we adopt a polytropic equation of state  
\begin{equation}  
\rm P = K \, \rho^{\gamma}
\label{polytr}
\end{equation}
where $\gamma$ is equal to 5/3 and K is a free parameter that we vary to produce different degrees of clumpiness. 
We consider 4 different values for K:  0.4665, 0.933, 1.3995 and 2.3325. As expected, the smaller K values produce a higher level of clumpiness, with denser condensations. For K values of 0.4665, 0.933, 1.3995 and 2.3325, the gaseous disk has respectively 60\%, 30\%, 20\% and 10\% of its total mass in regions with densities higher than $\rm 10^{5} \, cm^{-3}$.  


%

In the model that we just described, we introduce an MBH binary  in a circular orbit, and we follow the subsequent dynamical evolution of the system. We evolve the system using the SPH code called GADGET (Springel, Yoshida \& White 2001). We are interested in studying the evolution of the MBH binary separation with different values of the model parameters. The first parameter considered is the level of clumpiness of the gas, which we vary by varying the value of K. Secondly, since the total amount of gas in the nucleus of a merging system varies from one case to  another, and since the black hole masses may also vary, we consider several different values for the mass ratio between the MBHs and the gaseous disk. Thirdly, since the black holes need not orbit in the plane of the disk, we vary the inclination angle between the plane of the binary and the plane of the disk. Table 1 lists the parameters used in the different simulations.

\begin {center}
\centerline{Table 1: Run Parameters}
\begin{tabular}{cccc} \hline

RUN  &  $\rm M_{BH}/M_{gas}$  &  angle  &  K \\ \hline \hline A  &  0.01  & $0^{\circ}$  &  0.4665  
\\  B  &  0.01  & $0^{\circ}$  &  0.933 \\  C  &  0.01  & $0^{\circ}$  &  1.3995 \\  D  &  0.01  & $0^{\circ}$  &  2.3325 \\  E  &  0.01  & $22.5^{\circ}$  &   0.933 \\  F  &  0.01  & $45^{\circ}$  &  0.933 \\  G  &  0.01  & $67.5^{\circ}$  &  0.933 \\   H &  0.03  & $0^{\circ}$  &  0.933 \\  I  &  0.05  & 0$^{\circ}$  &  0.933 \\  J  &  0.1  & 0$^{\circ}$  &  0.933 \\  K  &  0.3  & 0$^{\circ}$  &  0.933 \\  L  &  0.5  & 0$^{\circ}$  &  0.933    

\\ \hline 

\end{tabular}
\end{center}


\section{Results}  

\subsection{Effects of Varying the Clumpiness}

We start by varying the level of clumpiness. We implement this by varying the constant K, considering  4 different values: 0.4665, 0.933, 1.3995 and 2.3325. To illustrate the effects, we consider first an MBH binary in which each black hole has a mass equal to  1\% of the total gas mass, initially in a circular orbit with a binary separation of 400 pc, or half of the disk diameter.

\begin{figure}
\plotone{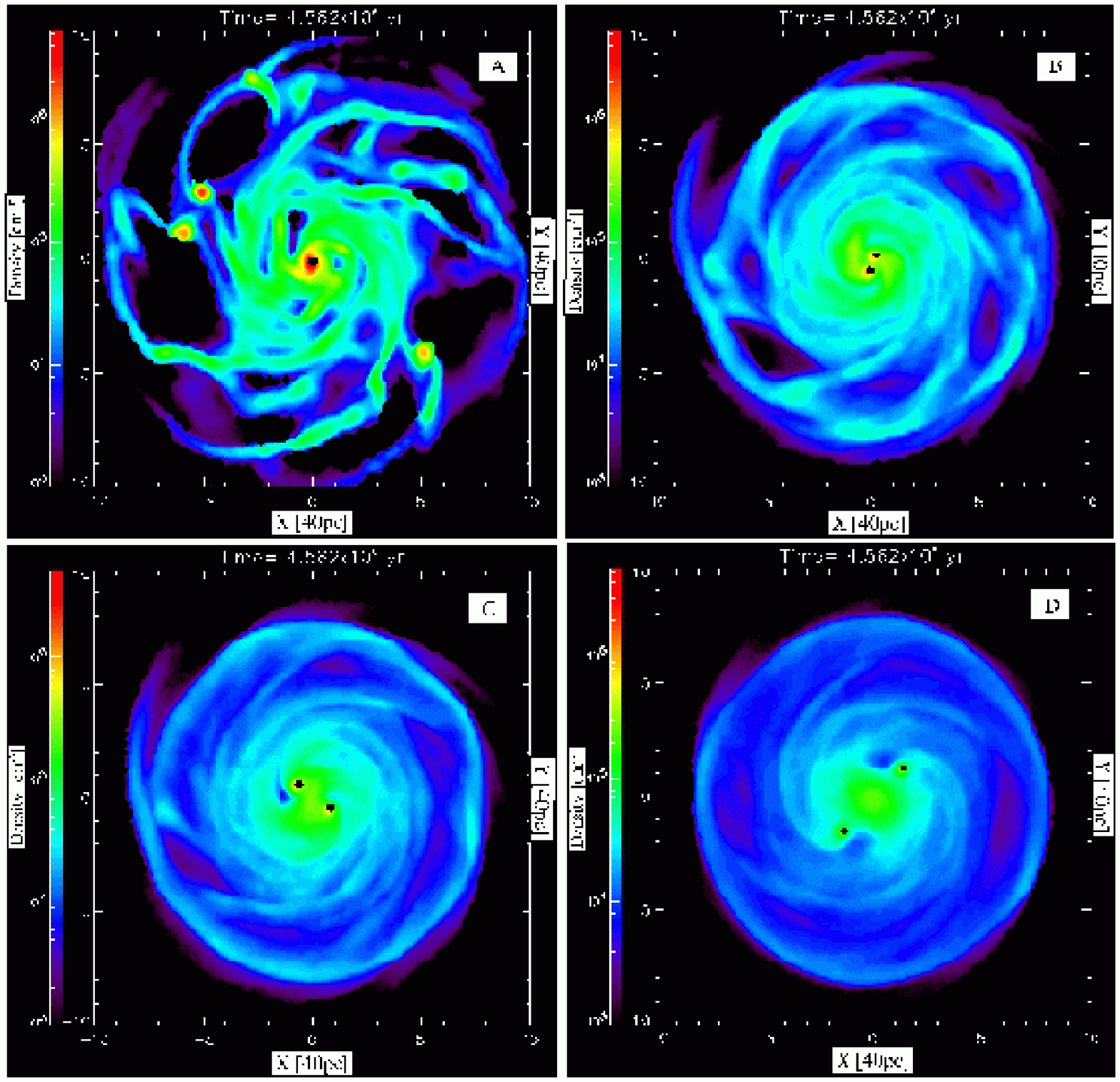}
\caption{Density distribution in the plane of the gas disk, coded on a logarithmic scale, at time $\rm t= 4.562 \times 10^{6} yr$  for  runs A, B, C and D, which have different assumed levels of clumpiness.  The  MBHs are   indicated in each plot by the black dots, and each MBH has 1\% of the mass of the gas disk. The clumpiness is varied by variying the parameter K in Eq. $\ref{polytr}$, as tabulated in Table 1, and the form of the resulting  density distribution varies from  relatively smooth spiral features for a large value of K (run D) to one with  dense clumps and filaments for small K (run A).  
\label{fig3}}
\end{figure}

Figure $\ref{fig3}$ illustrates a representative stage in the evolution of the system for the four different values of K, showing the face-on density distribution in the gas disk at the same  time in each case ($\rm t= 4.562 \times 10^{6} yr$). The density distribution varies from  relatively smooth spiral features to dense clumps and filaments as we reduce the parameter K. For K values of 0.4665, 0.933, 1.3995 and 2.3325, the disk thickness is respectively 20, 25, 32 and 40 pc. In our model, the disk thickness is determined by the sound speed $\rm c_{S}=(\frac{5}{3}K\rho^{2/3})^{1/2}$ implied by the adiabatic equation of state used, and that ranges from 62 to 78 $\rm  km \, s^{-1}$ for different values of K. In reality the sound speed $\rm c_{S}$ corresponds to internal turbulent motions in the gas, which are observed to be of this order, as noted in section 2. Our aim is to represent the observed highly turbulent molecular disk (Downes \& Solomon 1998) in which a turbulent velocity dispersion of $\rm \sim 100 \, km \, s^{-1}$ determines the disk thickness.

\begin{figure}
\plotone{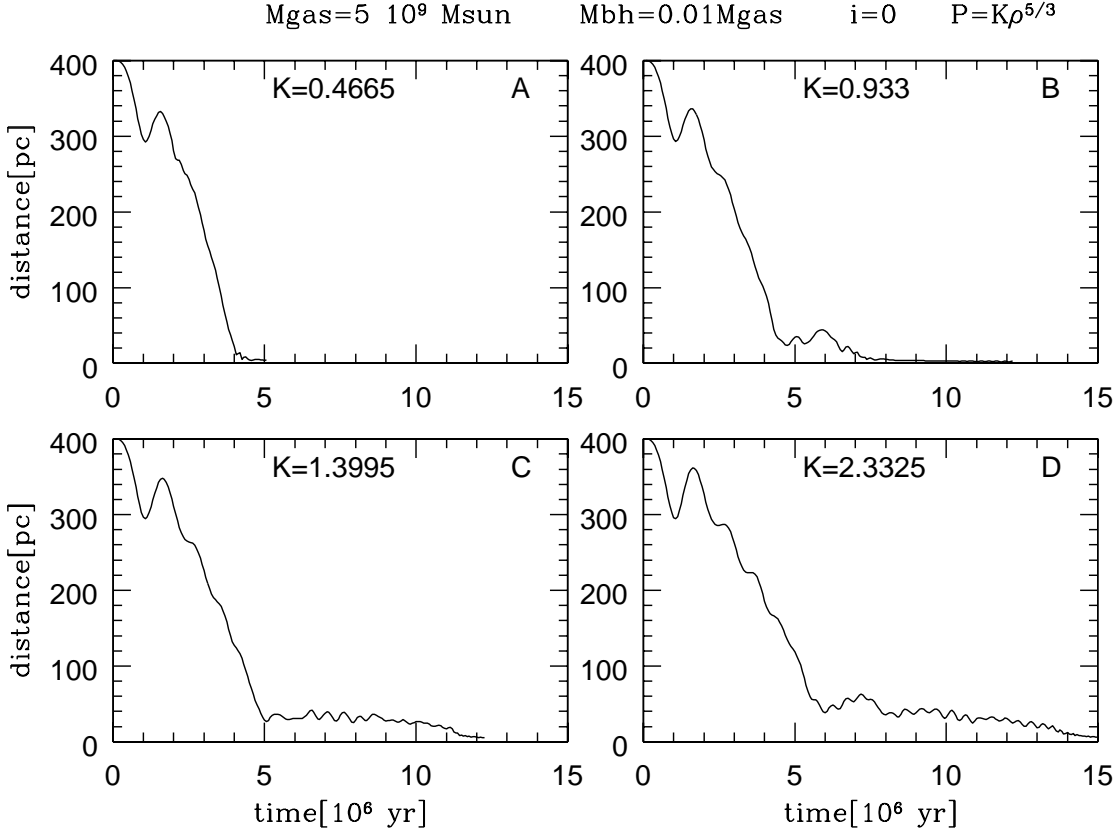}
\caption{This plot  shows the evolution of the binary separation for runs A, B, C and D with four different values of K, as indicated. In all cases  the mass of  each MBH is  1\% of the mass of the gas, and the orbit of the binary is in the plane of the disk. A transition phase with a period of slower evolution after $\rm t \sim  5 \times 10^{6} yr$ becomes increasingly prominent as K is increased and the gas disk becomes less clumpy.  
\label{fig4}}
\end{figure}

\begin{figure}
\plotone{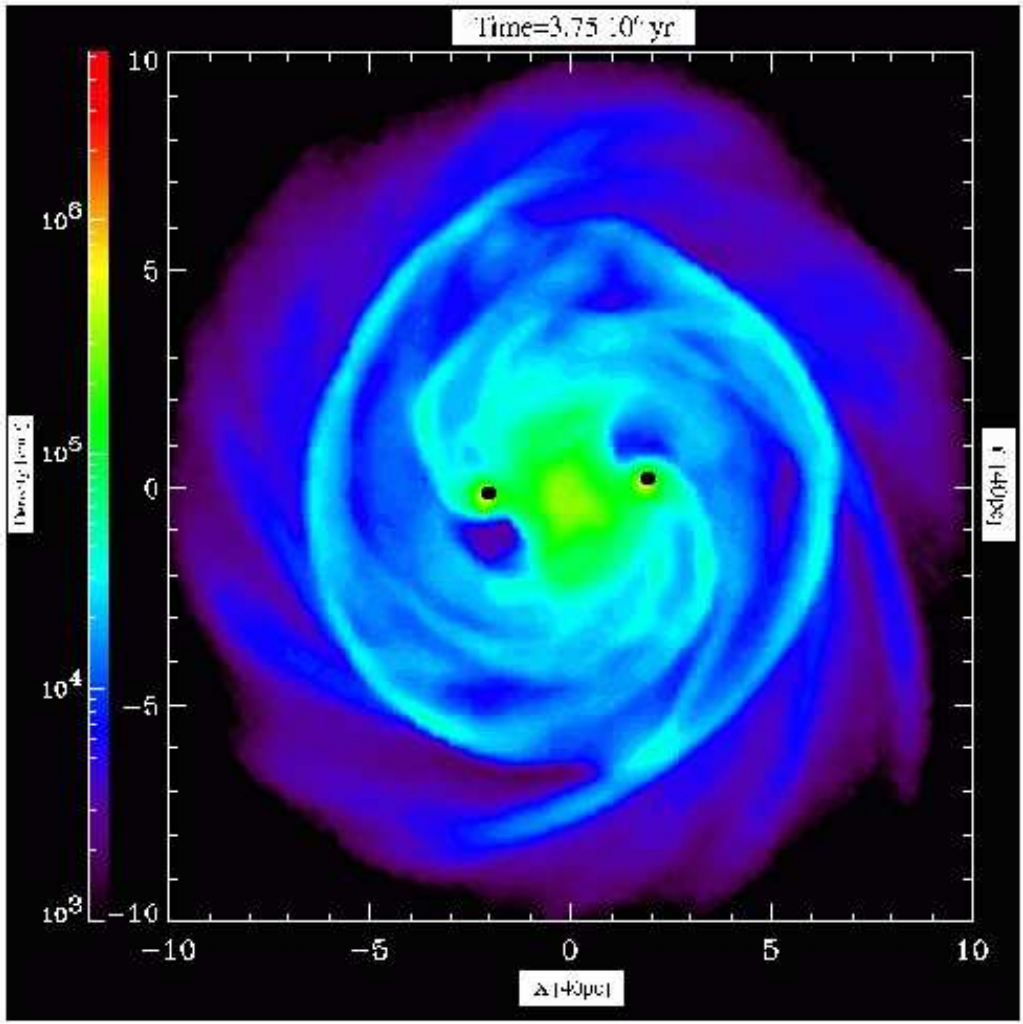}
\caption{Density distribution in the plane of the gas disk for run C at time $\rm t= 3.75 \times 10^{6} yr$, showing in more detail the spiral shocks produced by each orbiting MBH. Qualitatively as was found in Paper I, the response of the gas to the presence of the MBHs is a lagging density enhancement and a spiral shock that propagates outward from each MBH.
\label{fig4.1}}
\end{figure}

Figure $\ref{fig4}$ shows the evolution of the binary separation in these four cases over several orbits. In the early evolution of  the system, the separation diminishes due to the gravitational drag exerted by the background medium; this regime lasts until the binary separation is approximately 40 pc. Figure $\ref{fig4.1}$   shows the density distribution in the plane of the disk  at  time $\rm 3.75 \times 10^{6} yr$ for run C. As was found in Paper I, the response of the gaseous medium to the presence of the MBHs is a lagging density enhancement and a spiral shock that propagates outward from each MBH.

\begin{figure}
\plotone{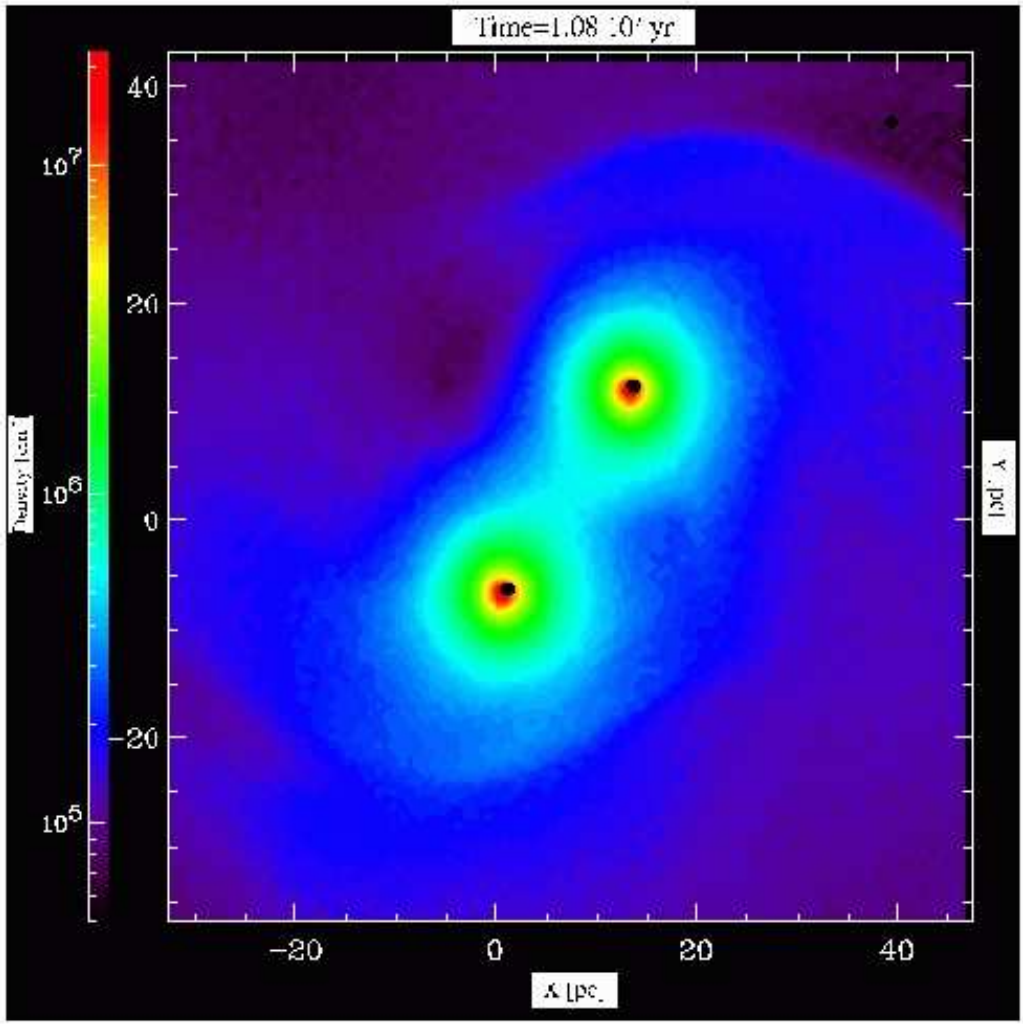}
\caption{Gas density  in the plane in the transition regime, illustrated for run C at time $\rm t= 1.08 \times 10^{7} yr$. The black dots  indicate as before the positions of the MBHs, each with 1\% of the mass of  gas. The figure indicates how the gravitational wake of each MBH is disrupted and smeared out by the gravity of the other MBH, accounting for the period of relatively slow evolution after $\rm t \sim  5 \times 10^{6} yr$ seen in Figure 2.
\label{fig4.2}}
\end{figure}

A notable feature of figure $\ref{fig4}$ is the marked increase of the coalescence timescale when the binary separation is less than 40 pc. This happens because the MBHs then come close enough that the wake formed by each MBH is disrupted by the gravity of the other MBH. As is shown in figure $\ref{fig4.2}$ for run C, the presence of the second MBH tends to smear out the wake, and therefore the gravitational drag decreases. This is not a surprising result because it is well known that, for both stellar and gaseous backgrounds, the dynamical friction approximation is no longer valid when the MBHs are at distances where the total mass of the background material within the orbit of the MBHs is comparable to the mass of the binary, because the background is then strongly perturbed by the presence of the MBHs. This physical regime corresponds to the transition between the gravitational drag and ellipsoidal torque regimes discussed in Paper I.

\begin{figure}
\plotone{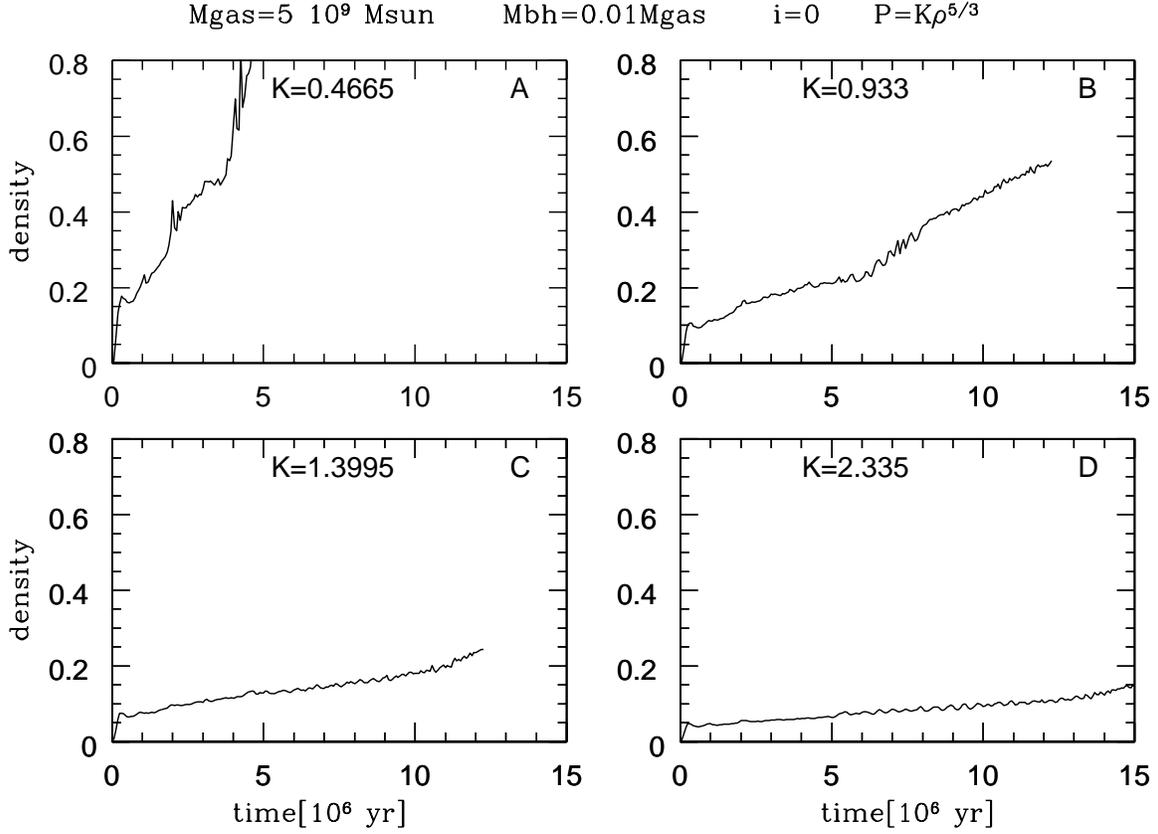}
\caption{This plot shows the average density within 20 pc around each MBH as a function of time for the 4 different values of K: 0.4665, 0.933, 1.3995 and 2.3325.  The density in the vicinity of each MBH increases strongly as K decreases, accounting for the faster evolution shown in Figure 2 for smaller values of K. 
\label{fig4.5}}
\end{figure}

The binary evolution has a strong dependence on the parameter K, and the  cases with lower K (higher clumpiness) suffer a stronger  deceleration. The explanation involves the gas density close to each MBH, which is higher for  lower K. As a result, because the gravitational drag is stronger in a denser environment, the binary MBH separation shrinks faster in the cases with lower K. The effect of K on the local density is illustrated in Figure $\ref{fig4.5}$, which shows the average gas density within 20 pc around each MBH,  as a function of time for the 4 different values of K. The figure shows that the density in the vicinity of each MBH varies strongly between the different cases, and  this produces important differences in coalescence timescale, as seen in  Figure $\ref{fig4}$. In run A, the drastic increase in density seen in Figure $\ref{fig4.5}$ compensates for the decrease in the efficiency of gravitational drag in the transition period, thereby maintaining  the coalescence timescale roughly constant. But as we increase K (respectively in runs B, C and D), the increase of density is less, making the increase in coalescence timescale more obvious. In the isothermal sphere model of Paper I, the gas becomes increasingly concentrated around the central MBH binary, qualitatively as in run A, and as a result there is no distinct transition period.

In section 5, we will show that in the later stages, the subsequent evolution of the binary leads to  rapid coalescence in only another $\rm \sim 10^{6} yr$. Therefore  the binary  spends   most of its time in the stages of evolution discussed above, making it important to understand the changes in coalescence timescale under different values of the model parameters.

\subsection{Effects of Varying the Orbital Inclination Angle}

Our second parameter is  the inclination angle between the plane of the disk and the plane of the binary. We perform simulations with four different inclination angles: 0, 22.5, 45 and 67.5 degrees. As in section 4.1, we consider an MBH binary initially in a circular orbit with MBH mass equal to the 1\% of the total gaseous mass, and for these runs we adopt an equation of state with K=0.933, as in run B. 

\begin{figure}
\plotone{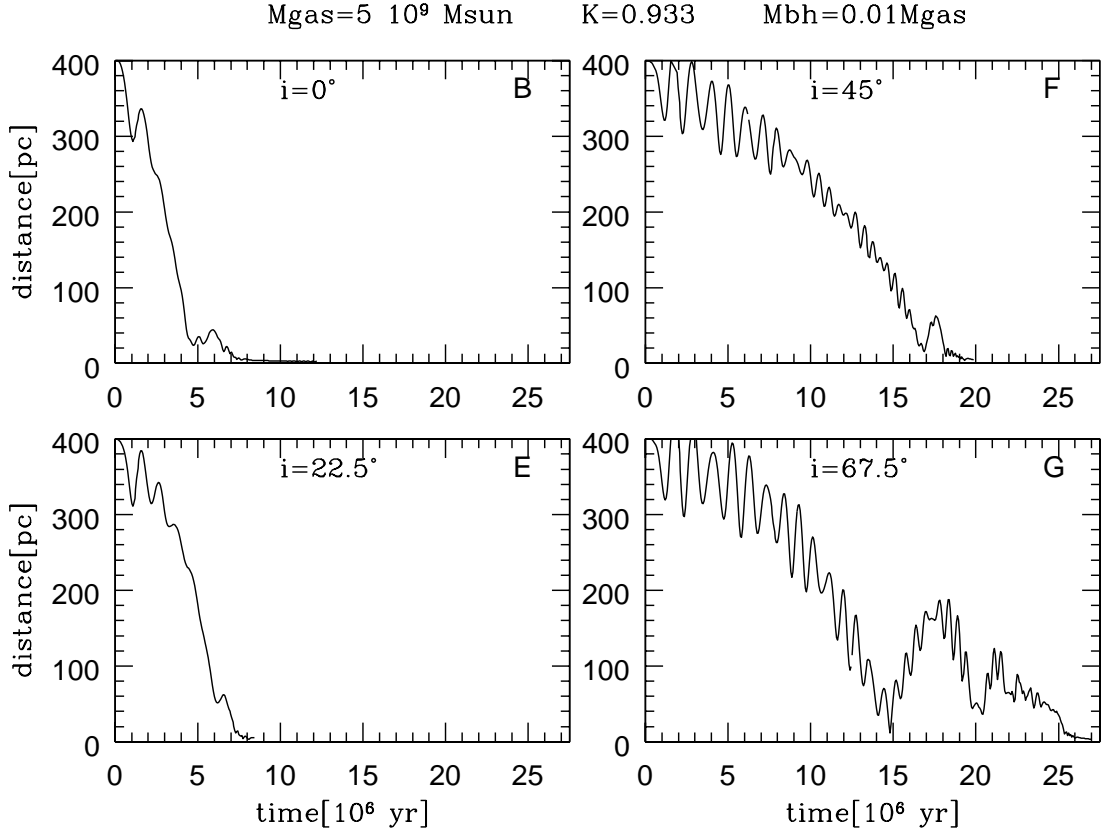}
\caption{This plot  shows the evolution of the binary separation for runs B, E, F and G with inclination angles of $\rm 0^{\circ}$, $\rm 22.5^{\circ}$, $\rm 45^{\circ}$ and $\rm 67.5^{\circ}$ between the plane of the disk and the plane of the binary. For all of these runs we adopt an equation of state with K=0.933 and the mass of each MBH is  1\% of the mass of the gas.
\label{fig6}}
\end{figure}

\begin{figure}
\plotone{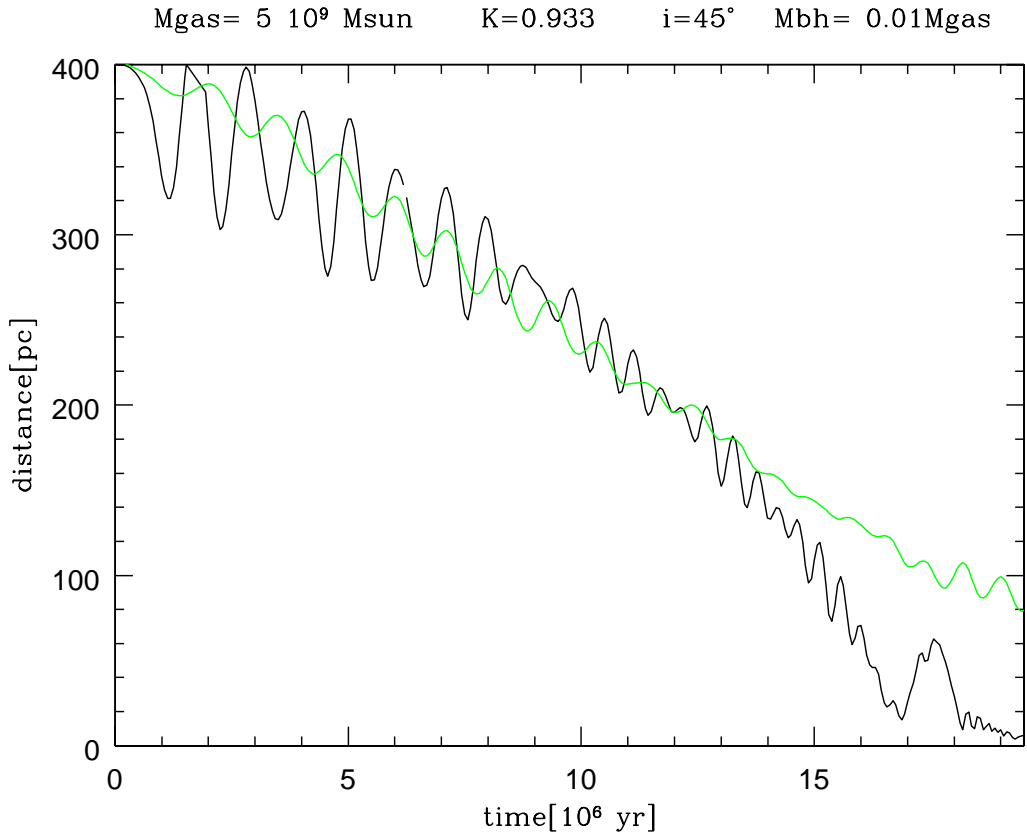}
\caption{This plot  shows the evolution of the binary separation for the case where  inclination angle  between the plane of the disk and the plane of the binary is $\rm 45^{\circ}$.  The black  curve shows an enlargement of the curve shown  for run F in Figure 6, and the green curve shows  the result of a run assuming the  same stellar bulge but no the gas disk.
\label{fig6.5}}
\end{figure}

Figure $\ref{fig6}$ shows the evolution of the binary separation in these four cases over several orbits. Although the case with i=$\rm 22.5^{\circ}$ has a coalescence timescale similar to the coplanar case, for  inclination angles of 45 and 67.5 degrees the coalescence timescales are increased by factors of 3 and 4 respectively. The reason for this difference is that in these cases, the early evolution of the binary is  driven by the gravitational drag exerted by the stellar bulge instead of by the gravitational drag of the gas  disk. Figure $\ref{fig6.5}$ shows a comparison of the evolution of the binary separation for the case with i=$\rm 45^{\circ}$ (black curve) with the evolution of the binary in the same bulge but without the gaseous disk (green curve). The agreement between these curves shows that the binary coalescence is driven up to time t=$\rm 1.25 \times 10^{7} yr$ by  the gravitational drag exerted by the stellar bulge. In all cases, the drag due to the gas disk eventually becomes dominant after the separation becomes small enough that the MBH binary spends more than 50\% of  its time in the disk, and the evolution thereafter proceeds much as in the coplanar case.

\subsection{Effects of Varying the Black Hole to Gas Mass Ratio}

Finally, we study the effect of varying the  ratio between the mass of the MBHs and the mass of the gas disk. We consider four cases with  black hole masses of 1\%, 3\%, 5\% \& 10\% of the total gas mass. We introduce a MBH binary initially in a circular orbit in the plane of the disk, using an adiabatic equation of state with K=0.933. 

\begin{figure}
\plotone{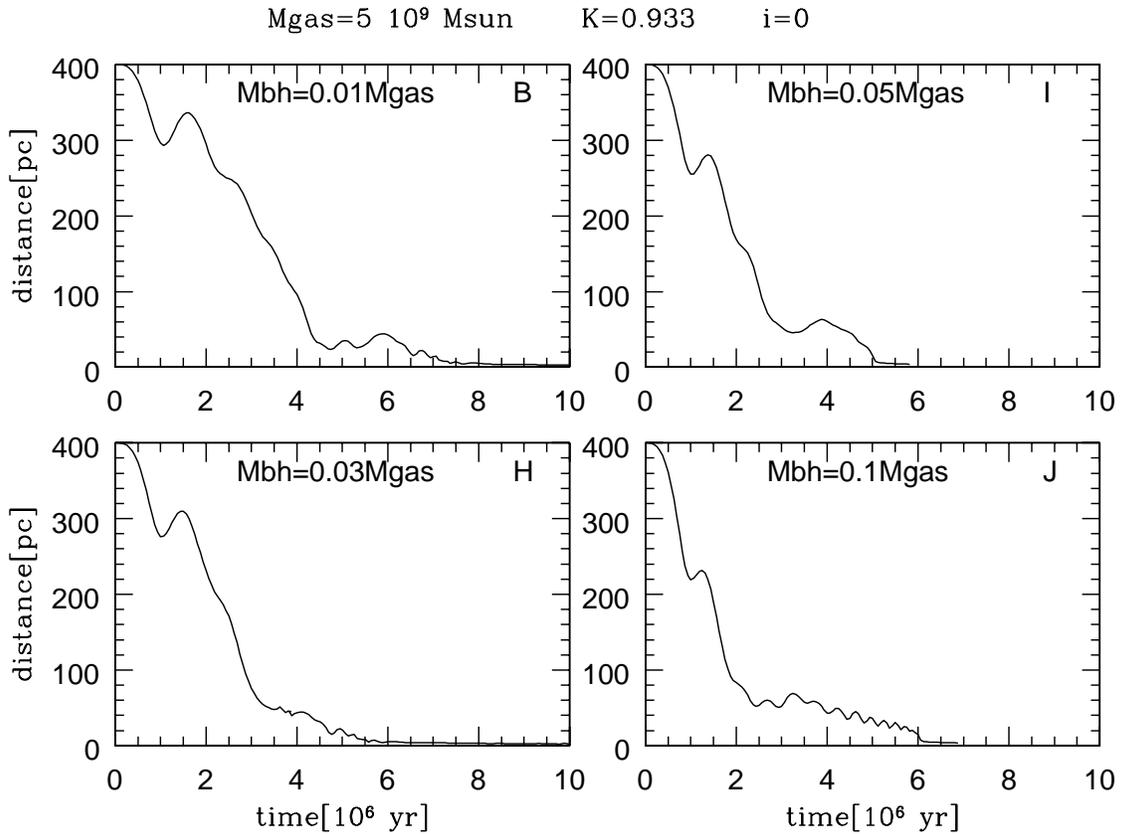}
\caption{This plot  shows the evolution of the binary separation for runs B, H, I and J which assume 4 different ratios between the mass of each MBH and the mass of the gas disk: 0.01, 0.03, 0.05, 0.1. For these runs we adopt an equation of state with K=0.933, and the  orbit of the binary is in the plane of the disk. Overall, there is only a weak dependence of the resulting coalescence time on the mass of the MBHs, owing mainly to the weak or even inverse dependence during the later stages.
\label{fig5}}
\end{figure}

Figure $\ref{fig5}$ shows the evolution of the binary separation in these four cases over several orbits.  In the early evolution of  the system, when the separation diminishes due to the gravitational drag exerted by the background medium, the binary coalescence timescale depends on the black hole mass, with the more massive cases suffering a stronger  deceleration. The time required for the binary separation to decrease by   75\% varies from  $\rm t \sim 4 \times 10^{6}$ yr for  $\rm M_{\rm BH}=0.01 \, M_{\rm gas}$  to  $\rm t \sim 2 \times 10^{6}$ yr for $\rm M_{BH} = 0.1 \, M_{\rm gas}$. In units of the  initial orbital period, this time varies from 2 to 1 orbital periods.  The trend is similar to that obtained in Paper I for a smooth gaseous medium, and is qualitativelly consistent with the Chandrasekhar formula.     
 
\begin{figure}
\plotone{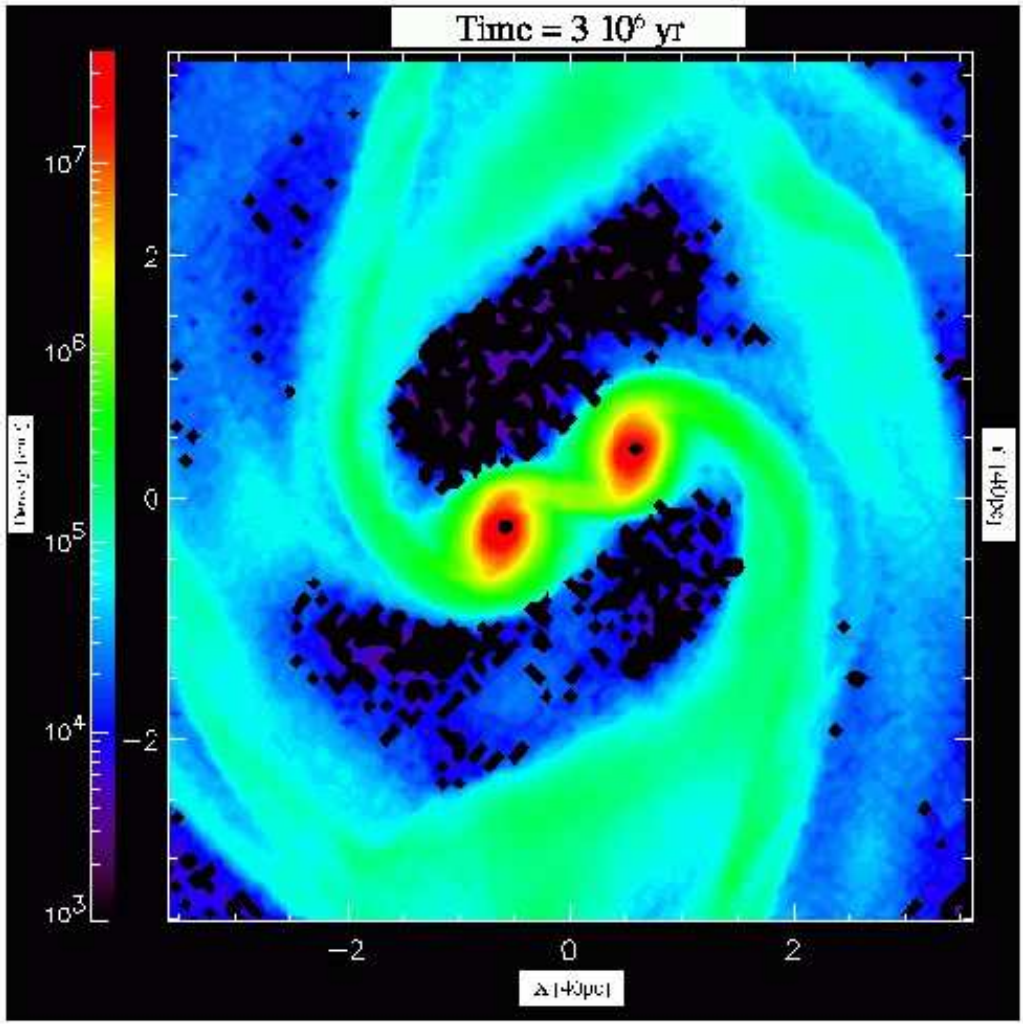}
\caption{The density distribution, in the plane of the disk  for  run J  at $\rm t= 3 \times 10^{6} yr$. The black dots  indicate the position of the MBHs and in this simulation the mass of  each MBH is  equal to 10\% of the mass of the gas. The gas disk is  globally  perturbed  and forms transitory gaps, as a response  to the presence of the MBHs.
\label{fig6.x}}
\end{figure}

The dependence of the binary coalescence timescale on the black hole mass changes when the binary separation arrives in the transition regime discussed in \S 4.1. Figure $\ref{fig5}$ shows that at later times there is no longer a clear dependence on the black hole mass; this is because as we increase the black hole mass, the wake formed is stronger, but the  disruption of the wake by the gravity of the other MBH is also stronger, weakening the strong correlation with  black hole mass. In fact, for the case of  $\rm M_{BH} = 0.1 \, M_{\rm gas}$, we see a small  $increase$ in the coalesce timescale in the transition regime. In this case the disk is starting to be globally  perturbed by the presence of the MBHs as seen in Figure $\ref{fig6.x}$, and not only locally as before. Tidal and/or resonant forces are starting to evacuate gas from the central region, and these effects become more important as we continue increasing the black hole mass. As a result, in the overall evolution of the binary separation, the dependence  on the black hole mass is only weak, mainly because there is not a clear dependence on mass during the later times. 

\begin{figure}
\plotone{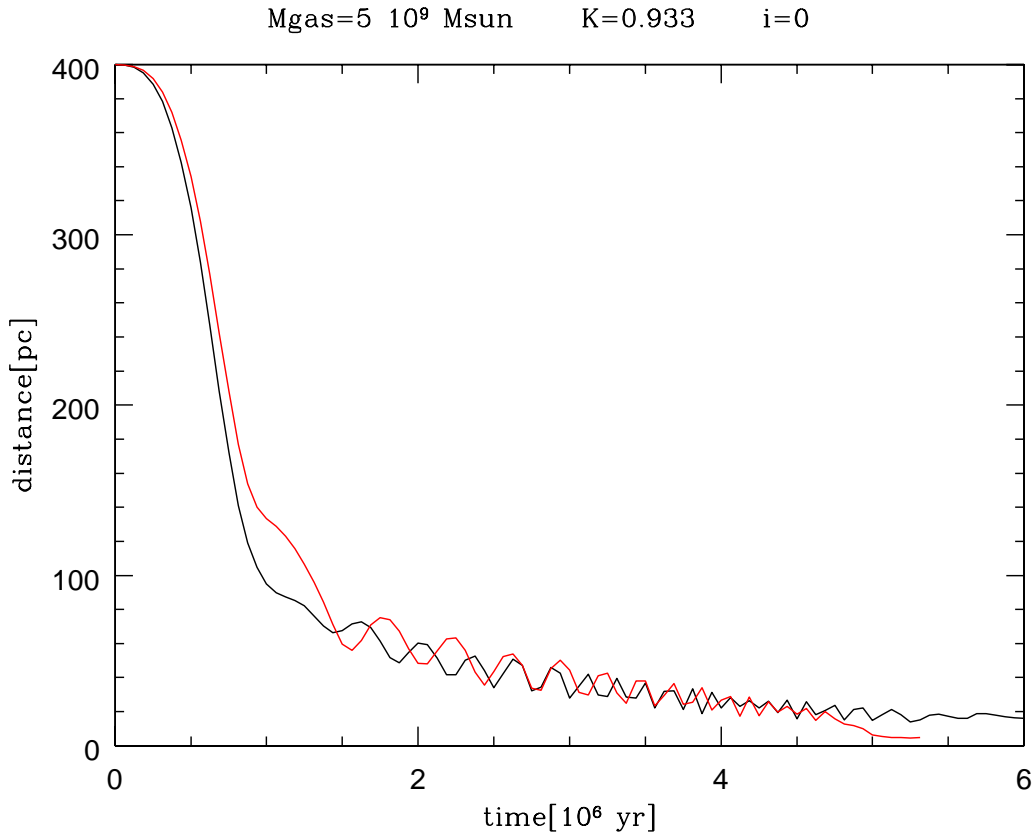}
\caption{This plot  shows the evolution of the binary separation for  runs K and L which assume relatively large MBH masses.  The  black curve shows the evolution of the binary when $M_{\rm BH} = 0.5M_{\rm gas}$, and the red  curve shows the  calculation with $M_{\rm BH} = 0.3M_{\rm gas}$. For these runs we adopt an equation of state with K=0.933, and the binary's orbit is in the plane of the disk. The evolution of the run with  $M_{\rm BH} = 0.5M_{\rm gas}$ almost stalls at late times because of the dearing of a gap in the disk, as illustrated in Figure 10. 
\label{fig6.6}}
\end{figure}

In order to determine when the gas becomes no longer important in driving the coalescence of the binary, we continue increasing the black hole mass. We consider two more  cases with  black hole masses of 30\% and 50\% of the total gas mass. Figure $\ref{fig6.6}$  shows the evolution of the binary separation in these two cases over several orbits. We found that the coalescence tends to stall in the case where the black hole mass equals 50\% of the total mass (red line in Figure $\ref{fig6.6}$). This happens because strong tidal and/or resonant forces create a circumbinary gap, as seen in Figure $\ref{fig6.7}$, and therefore the gravitational drag is no longer effective. The interaction of a binary with a circumbinary disk in cases where the total gas mass is typically equal to or smaller than the binary mass has been widely studied in the context of star/planetary formation (Lin \& Papaloizou 1979; Goldreich \& Tremaine 1980; Artymowicz \& Lubow 1994,1996; Armitage \& Natarajan 2002), where a similar circumbinary gap is found. The criterion for opening a gap will be discussed in \S 6.

\section{Final Evolution Using Higher Resolution Simulations}   

The runs described above were continued until the MBH separation approaches the assumed gravitational softening length of 4 pc, at which point the evolution of the system artificially stalls. To continue the evolution of the binary it is necessary to reduce the gravitational softening length, something that is extremely expensive computationally. We choose to reduce  it from $\epsilon_{\rm soft}$=4 pc  to 0.1 pc in order to follow the evolution of the binary separation by one more order of magnitude, and we apply this  procedure to the cases  where  each MBH has 1\% of the total mass in gas and the binary is in the plane of the disk, again for the 4 runs with different values of K. We reduce the softening length when the binary sepatation is approximately  30 pc, which occurs at different times  for different values of K.

Figure $\ref{fig7}$ shows the evolution of the binary separation in these four cases over several orbits.  The black curves are the same  calculations shown in Fig. $\ref{fig4}$  but on a logarithmic scale, and the red curves show the results for the simulations with smaller softening length. 

In the early evolution of the system, the black and red curves show almost the same behavior in each case. The situation changes drastically  when the binary separation becomes less than about 10 pc, that is, when the binary arrives at separations comparable to the 'gravitational influence' radius of the black hole: $\rm R_{inf}=2GM_{BH}/(v^{2}_{BH}+c^{2}_{S})$. For example, at time $\rm 4 \times 10^{6} yr$ in run A, in which the black hole mass $\rm M_{BH}$ is $\rm 5\times 10^{7}\, \msun$, the MBH velocity $\rm v_{BH}$ is $\rm 172 \, km\, s^{-1}$ and the sound speed $\rm c_{S}$ is $\rm  (\frac{5\, P}{3\, \rho})^{1/2}=158 \, km \, s^{-1}$  in the vicinity (within 5 pc) of each MBH, the gravitational influence radius $\rm R_{inf}$ is 7.9 pc. At these distances the binary completely dominates  the gravitational potential in its vicinity, and  the response of the  medium to that gravitational field is  a trailing ellipsoidal density enhancement, as was found in Paper I. The formation of the ellipsoidal density enhancement typically occurs when the binary separation is about  $\rm 1.5 \, R_{inf}$, and that value ranges from 9 to 12 pc in these four cases. Therefore the transition to the new regime is only weakly dependent on K.
 
Figure  $\ref{fig7.5}$ shows the density enhancement produced by the MBH's in the gaseous medium, for the simulation with K=1.3995  at four different times: 10.8, 11, 11.2 and $\rm 11.4 \times 10^{6} yr$. Figure $\ref{fig7.5}$  clearly shows that the rapid decay in the binary separation, which starts at around t=$\rm 11.2 \times 10^{6} yr$, coincides with the formation of the ellipsoid. In this figure, the black circles centered on each MBH indicate the gravitational influence radius $\rm R_{inf}$, and the figure shows that the trailing ellipsoid forms when these circles overlap. The axis of the ellipsoid is not coincident with the binary axis but lags behind it, and this  offset  produces a torque on the binary that is now  responsible for the angular momentum loss. This mechanism was extensively studied in Paper I and is able to reduce the binary separation to distances where gravitational radiation is efficient.

In the simulations described above, we reduced the gravitational softening length but we maintained the number of SPH particles used. In order to verify that we have enough resolution to resolve adequately the region inside r = 40 pc, where the dynamics of the final coalescence occurs, we repeat one of these simulations increasing the number of particles. We use  the particle splitting procedure described in Paper I to achieve this goal. The  procedure is applied to the case  where  K=1.3995. We split the SPH particles at a  time of $\rm  10^{7} yr$ when the binary enters  the r $<$ 30pc region; for  each parent particle we introduce $N_{\rm split}$=8 child particles and therefore we increase the number of particles to 1,882,648.  

Figure $\ref{fig8}$ shows the evolution of the binary separation for the high-resolution calculation. The red curve is the same  calculation shown in Fig. $\ref{fig7}$, and the green curve shows the MBH's separation in the high-resolution calculation. The result is qualitatively the same and quantitatively very similar to the low-resolution calculation. This supports the validity of the results shown by the red curves in Fig. $\ref{fig7}$, based on the low-resolution calculations.

\subsection{Gravitational Radiation}

The final phase in the evolution of a binary MBH occurs when eventually the MBHs  become close enough to allow  gravitational radiation to become an efficient mechanism for angular momentum loss.

For an equal mass binary and scaled to convenient units, the gravitational radiation timescale (Peters 1964; see Eq. 16 in Paper I) is:
\begin{equation}
t_{\rm gr} = 2.9 \times 10^{6}{\rm yr} \, \left(\frac{a}{\rm 0.01 pc}\right)^{4} \left(\frac{\rm 10^{8}\msun}{\rm M_{BH}}\right)^{3}  F(e) \, ,
\label{quinlan}
\end{equation}
where $a$ is the binary separation, $\rm M_{BH}$ is the mass of each black hole, and $F(e)$ (see Paper I) contains the eccentricity dependence, which is weak  for small $e$; e.g. $F$(0)=1 and $F$(0.5)$\sim$0.205. For a binary in which the mass of each MBH is $\rm 5 \times 10^{7}\msun$ (corresponding in our simulations to the case where each  MBH has 1\% of the total mass in gas) and which has a separation of 0.01 pc, the gravitational radiation coalescence timescale is ${ t}_{\rm gr} = 1.16 \times 10^{7}$yr. If we extrapolate our results  to  separations of 0.01pc, we predict that the MBH binary will merge in  $\sim 10^{7}$yr for all values of K. We feel confident that we can  extrapolate our results to smaller separations because the ellipsoidal torque model studied in Paper I describes successfully the late evolution of the binary MBH, and it predicts that the separation should reach 0.01 pc within less than another $10^{6}$yr. 

\section{Gap Opening Criteria}

As we found in \S 4.3, when the MBH binary is able to open a circumbinary gap in the  disk, the gas no longer plays an important role in  the coalescence of the binary; the problem then reduces to the  evolution of a black hole binary at the center of a stellar system, which has been extensively studied in the literature  (Begelman, Blandford, \& Rees 1980; Makino \& Ebisuzaki 1996; Quinlan 1996; Milosavljevic \& Merritt 2001, 2003). Therefore it is important to determine the critical MBH mass needed to open the gap.

\begin{figure}
\plotone{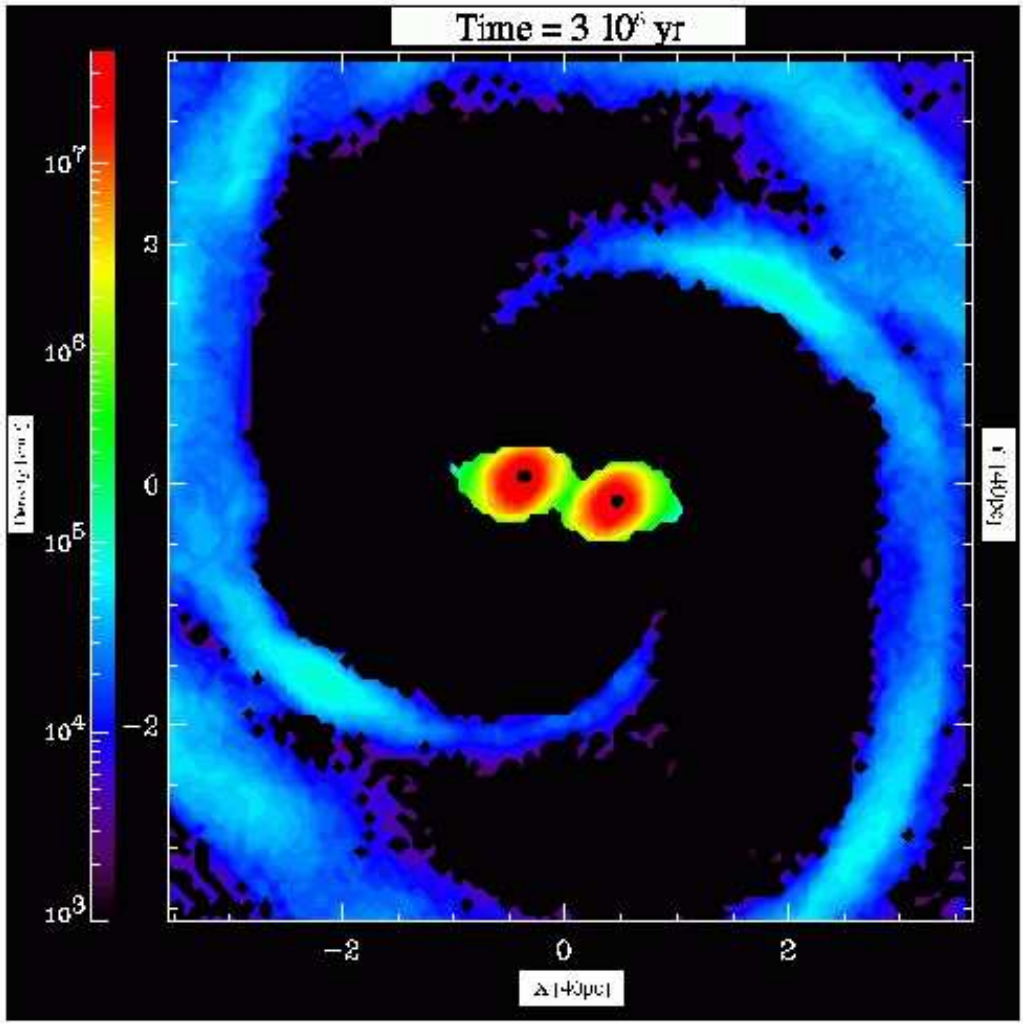}
\caption{The density distribution, in the plane of the disk  for  run L  at $\rm t= 3 \times 10^{6} yr$. The black dots  indicate the position of the MBHs and in this simulation the mass of  the binary is  equal to the mass of the gas. The gas response  to the presence of the MBHs is  a circumbinary gap created by the strong tidal and/or resonant forces near the MBHs.
\label{fig6.7}}
\end{figure}

\begin{figure}
\plotone{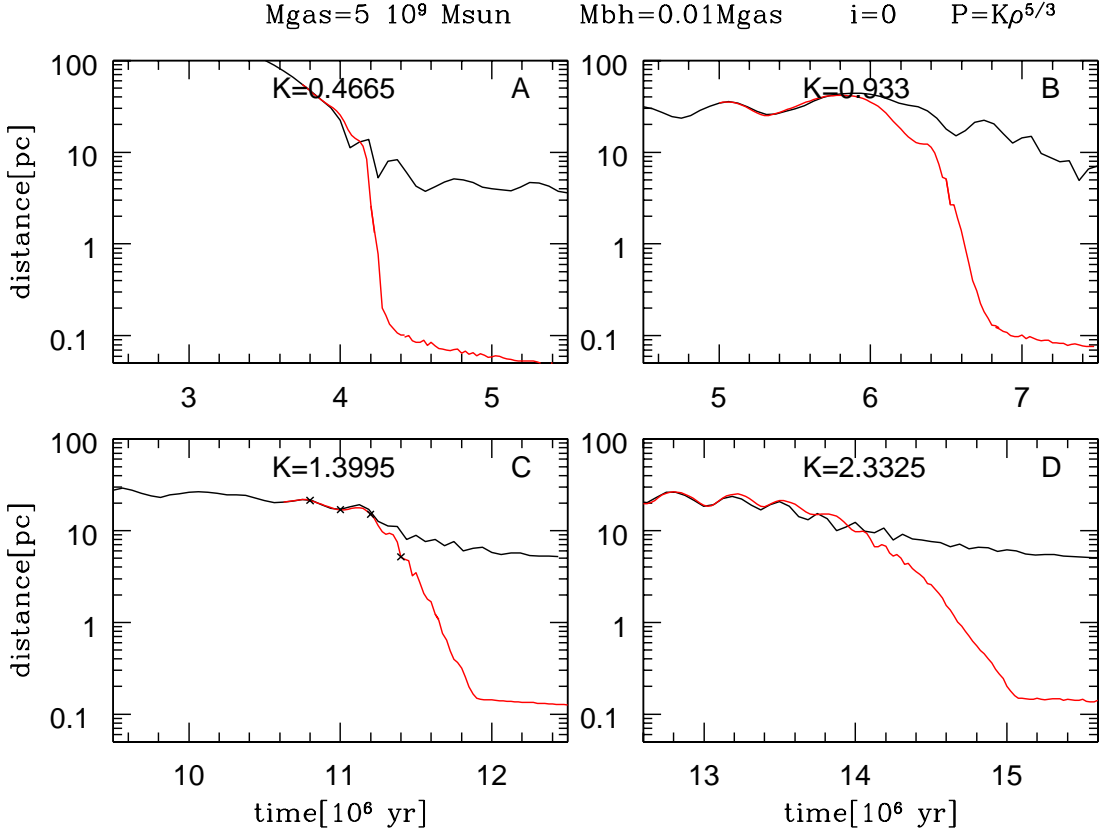}
\caption{This plot  shows the final evolution of the binary separation for runs A, B, C and D with four different values of K. In all cases the mass of each MBH is  1\% of the mass of the gas and the orbit of the binary  is in the plane of the disk. The black curves are the same calculations shown in Fig. $\ref{fig4}$ but  on a logarithmic scale. The red curves show the results for the simulations with smaller softening length. The black crosses in run C correspond to the four different figures shown in Fig. $\ref{fig7.5}$.
\label{fig7}}
\end{figure}

\begin{figure}
\plotone{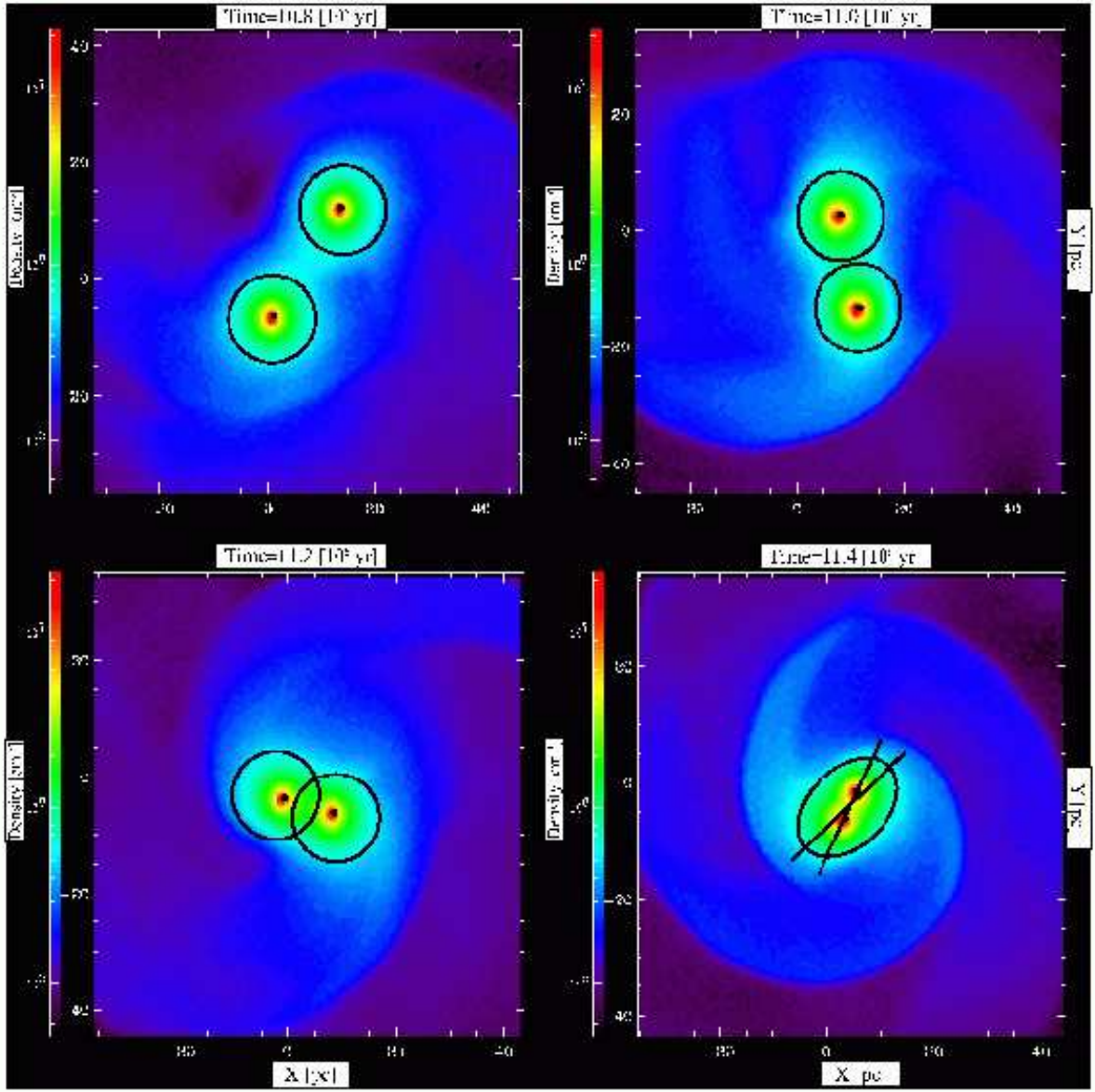}
\caption{Density distribution in the plane of the gas disk for run C, illustrated  at four different times: 10.8, 11, 11.2 and $\rm 11.4 \times 10^{6} yr$ that are indicated by crosses in Fig. $\ref{fig7}$c. When the regions within the gravitational influence radius of each MBH (black circles in the figure) merge at around $\rm 11.2 \times 10^{6} yr$, the gas forms a trailing ellipsoidal density enhancement bound to the binary that is  responsible for the rapid decay in the binary separation at late times that is seen in Fig. $\ref{fig7}$.
\label{fig7.5}}
\end{figure}

\begin{figure}
\plotone{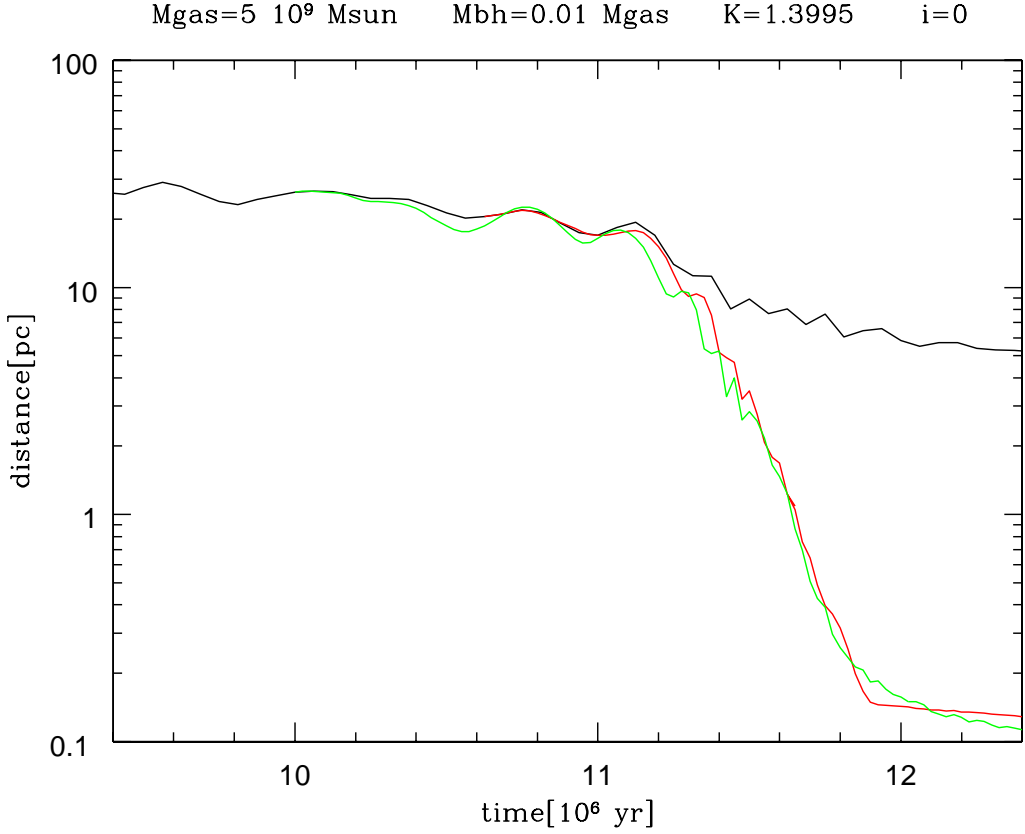}
\caption{This plot  shows the final evolution of the binary separation for the case where K=1.3995 (run C). The black curve is the same calculation shown in Fig. $\ref{fig4}$ but  on a logarithmic scale. The red curve shows the result for the simulation with smaller softening length, the same  calculation shown in Fig. $\ref{fig7}c$. The green curve shows the MBH's separation in the calculation that has both a smaller softening length and a much higher numerical resolution ($N_{SPH}=$1,882,648). The result is qualitatively the same and quantitatively very similar to the low-resolution calculation. This supports the validity of the results shown by the red curves in Fig. $\ref{fig7}$, based on the low-resolution calculations.
\label{fig8}}
\end{figure}

The criterion for opening a gap can be determined by comparing the gap-closing and gap-opening times (Goldreich and Tremaine 1980). The angular momentum $\Delta L$ that must be added to the gas to open a gap of radius $\Delta r$ in a disk with thickness h is $\Delta L \approx \rho (\Delta r)^{2} h r v$. Using the dynamical friction formula (Chandrasekhar 1943; Ostriker 1999), we infer that the torque that each black hole exerts on  the disk is 
\begin{equation}
T \approx 4\pi \rho r \left(\frac{G M_{\rm BH}}{v_{\rm BH}}\right)^{2} \times f^{\rm (star,gas)}\left({\cal M} \right)  
\end{equation}
where  ${\cal M}$=$v_{\rm BH}$/$C_{\rm S}$ and $f^{\rm (star,gas)}$ is a dimensionless factor which depends on the nature, stellar or gaseous, of the background medium (see Eqs. 7 to 10 in Paper I). This torque will supply angular momentum $\Delta L$ in a time $\Delta t_{\rm open} = \Delta L/T$. The tendency of for gap formation is opposed by turbulent diffusion, which fills up a gap of width $\Delta r$ on a timescale $\Delta t_{\rm close} = \Delta r/v_{\rm turb}$; as mentioned before, in our  adiabatic model the turbulent velocity dispersion is represented by the sound speed $(C_{\rm S}=v_{\rm turb})$. Gap formation occurs if $\Delta t_{\rm close} \gtrsim \Delta t_{\rm open}$, and taking into account that $v_{\rm BH} = (G M(r)/r)^{1/2}$ where M(r) is the total enclosed mass at  radius r, this criterion can be written    
\begin{equation}
M_{\rm BH} \gtrsim \sqrt{\frac{\alpha}{4\pi f({\cal M}) {\cal M}}} \, \left( \frac{h}{r}  \right) M(r) \, ,
\label{mhole}
\end{equation}
where $\alpha$ is the ratio of gap radius to disk  thickness, defined  by $\Delta r = \alpha h$. The simulation with  black hole masses equal to 50\% of the total gas mass suggests that a stable circumbinary gap forms when the MBHs are at a radius $ r \sim h = {\rm 25pc}$, with a gap width of $\Delta r \sim 3h = {\rm 80pc}$ $(\alpha = 3)$. For the parameters used in our simulation, Equation $\ref{mhole}$ corresponds to the condition $\rm M_{BH} \geq 0.46 = 0.46 \, M_{gas}$, in agreement with our result of Section 4.3 (Figure $\ref{fig6.7}$) that a cicumbinary gap opens when $\rm M_{BH} = 0.5M_{gas}$.

As seen in Equation $\ref{mhole}$, the critical mass at a given radius depends on the total enclosed mass. Because  approximately 80\% of the dynamical mass in the region of interest  is  contributed by the stellar bulge (Downes and Solomon 1998), we will assume that $\rm M(r) = G^{-1} r \sigma_{c}^{2}$,  $\sigma_{c}$ being the central stellar velocity dispersion. The  mass of the central black holes in nearby galaxies is observed to be correlated with the central velocity dispersion, measured within the central kpc, by the so-called `$m-\sigma_{c}$' relation (Tremaine et al 2002; Merritt and Ferrarese 2002):
\begin{equation}
M_{\rm binary} = {\rm 2} M_{\rm BH} = \left( \frac{\sigma_{c}}{\rm 200 \, km \,s^{-1}}\right)^{4} \, 10^{8} \msun \, ,
\end{equation}
where we assume that the  mass of the binary correlates in the same way with the central velocity dispersion of the merged bulges. Using this relation, Equation $\ref{mhole}$ can be written in a simpler form:
\begin{equation}
M_{\rm BH} \gtrsim \frac{\alpha}{4\pi f({\cal M}) {\cal M}} \left( \frac{h}{\rm pc}\right)^{2} \, 1.74 \cdot 10^{6} \msun \, .
\end{equation}
As mentioned earlier, our result of Section 4.3 (Figure $\ref{fig6.7}$) suggests that a stable circumbinary gap forms when  $\Delta r \sim 3h$, and therefore  we set $\alpha = 3$. For a Mach number of the order of unity, the critical mass required for a gap to form then reduces to
\begin{equation}
M_{\rm BH} \gtrsim  \left( \frac{h}{\rm pc} \right)^{2} \, 7.2 \cdot 10^{5} \msun \, .
\end{equation}
For typical  ULIRGs (Downes and Solomon 1998), h is at least of the order of 40pc. Therefore the formation of a circumbinary gap will be prevented, since  the  observed MBH masses range from $10^{6}$ to a few times $10^9$~M$_\odot$ and thus are smaller than the critical mass required to open a gap as given by Equation 9. The formation of a circumbinary gap will be prevented and the MBH binary will therefore coalesce as long as the disk thickness remains of the order of tens of  parsecs. Observations show that even in nuclear molecular disks of mass $\sim 10^{8} \msun$ the velocity dispersions are still $\rm \sim 40 \, km \, s^{-1}$, indicating that disk thicknesses of the order of tens of  parsecs are present, and this result is independent of whether or not the galaxy hosts a starburst (Jogee, Scotville and Kenney 2004). So gas should play a dominant role in driving the mergers of black holes as long as the amount of gas present in the central kpc is equal to or larger than this. In most  galaxy mergers it is expected to have a nuclear disk of at least this mass, because a disk mass of the order of $10^{8}\msun$  corresponds for an average spiral galaxy like the Milky Way to only 0.1\% of the  total mass being in gas, a small gas mass fraction compared to the 1 to 50\% present in Sa to Scd galaxies (Young et. al. 1995). Taking into account that in a galaxy merger  typically 60\% of the gas originally present in the merging galaxies  ends up in a massive nuclear disk, and  allowing for some gas depletion due to star formation, we can conservately conclude that in a merger of galaxies with at least 1\% of their total mass in gas, a nuclear disk with mass ${\rm \geq 10^{8}}\msun$ will be present, and that the gas will then play a major role in the evolution and  final coalescence of the binary.

On the other hand, if the masses of the black holes are greater by an order of magnitude or more than is expected from the `$m-\sigma_{c}$' relation, it is more probable that the MBH binary will form a circumbinary gap inside the central 100pc of the disk. In this case, the feeding of gas  into the MBHs and therefore their growth will be almost stopped. This suppresses an important accretion phase for MBHs, in which a non-negible part  of the MBH mass is expected to be assembled  because without the gap, the accretion is predicted to be at a super-Eddington rate (Dopita 1997). However the bulge will continue accreting  gas and growing, especially  in the central kiloparsec where $\sigma_{c}$ is measured, and therefore the stellar velocity dispersion will  also continue growing and trying to reach the `$m-\sigma_{c}$' relation. Thus gap formation can act as a self-regulatory mechanism for MBH growth, and this may help to explain the existence of the `$m-\sigma_{c}$' relation.

\section{Conclusions}

In our previous paper (Paper I) we studied the role of gas in driving the evolution of a binary MBH, and we followed its evolution through many orbits and close to the point where gravitational radiation becomes important. In Paper I, we presented results for a relatively idealized  case in which the gas was assumed to be in a nearly spherical and relatively smooth distribution, and we found important   differences in the evolution of a binary MBH in a gaseous medium,  compared to a stellar background. In particular, we didn't find any sign of ejection of the surrounding gas, as happens with stars in the later stages in the evolution of a binary MBH in a stellar system.

In the present paper, we have extended this work by studying more realistic models in which the gas is in a disk with significant clumpiness. We do not attempt a fully realistic description of the gas in merging galaxies, which often exhibits starburst activity, but adopt a simple polytropic equation of state with a variable coefficient that allows us to represent gas with a varying degree of clumpiness. In this way we are able to represent gas with the observed range of densities in merging systems. We also vary the inclination angle between the plane of the binary and the plane of the disk, and the mass ratio between the MBHs and the gaseous disk.

In the early evolution of  the system, as in Paper I, the separation diminishes due to the gravitational drag exerted by the background medium. We find a new  transition regime that was not apparent in Paper I, when the MBHs come close enough that the wake formed by each MBH is disrupted by the gravity of the other MBH. In this transition regime the gravitational drag is less effective and therefore the coalescence timescale increases. 

In the variety of simulations that we perform,  we find  that varying the level of clumpiness of the gas, or the  inclination angle between the plane of the binary and the plane of the disk, changes the binary coalescence timescale by less than a factor of 3. We also find that the dependence  on  the mass ratio between the MBHs and the gaseous disk  is only weak, mainly because there is not a clear mass dependence in the transition regime. In all of the cases considered, the coalescence timescale varies between $\rm 5\times 10^{6} yr$ and $\rm 2.5\times 10^{7} yr$, or between 2.5 and 12 initial orbital periods.
 
The final evolution  of the binary, when  the binary MBH dominates the gravitational potential in its vicinity, is driven by the same trailing ellipsoidal density enhancement that was found in Paper I. The axis of the ellipsoid lags behind  the binary axis, and this  offset  produces a torque on the binary that is now  responsible for the continuing loss of angular momentum. The formation of the ellipsoidal density enhancement typically occurs when the binary separation becomes comparable to the 'gravitational influence' radius of the black hole: $\rm R_{inf}=2GM_{BH}/(v^{2}_{BH}+c^{2}_{S})$. As was studied in Paper I, this mechanism is able to reduce the binary separation to distances where gravitational radiation is efficient and will cause a merger within another  $10^{7}$yr. 

We found that gas plays an important role in the coalescence of the binary when the MBHs  satisfy the observed `$m-\sigma_{c}$' relation and the disk thickness is of the order of tens of  parsecs. This  disk thickness is inferred from observations whenever gas is present in amounts greater than $\rm 10^{8} \msun$; this  corresponds, for an average spiral galaxy like the Milky Way, to only 0.1\% of the total mass being in gas. We predict that the MBH binary will merge within a few times $10^{7}$yr for the various values of the model parameters studied. Galaxies typically merge in  $10^{8}$ yrs, and therefore these results imply that in a merger of  galaxies with at least 1\% of their total mass in gas, the MBHs will coalesce soon after the galaxies merge. We also found that only if the MBHs considerably depart from the `$m-\sigma_{c}$' relation is it probable that the MBH binary will form a circumbinary gap that stalls the coalescence, but this gap formation can act as a self-regulatory mechanism that helps  to explain the existence of the `$m-\sigma_{c}$' relation.

The final coalescence has crucial implications for possible scenarios of massive black hole evolution and growth. Our work supports scenarios of hierarchical build-up of massive black holes, that in their most recent versions (Kauffmann \& Haehnelt 2000; Haehnelt 2003a; Volonteri, Haardt \& Madau 2003; Di Matteo et al. 2003) assume that the first `seed' black holes appear at high redshifts (z $>$ 10) with intermediate masses ($\rm \sim 300 \msun$), and that the black holes grow by mergers and gas accretion following the merger hierarchy from early times until the present. In particular, this result supports the assumption that halo mergers lead to black hole mergers when gas is present, something probably always true at high redshift. The final coalescence of the black holes leads to gravitational radiation emission that would be detectable up to high redshift by LISA (Menou 2003 or Haehnelt 2003b). This is  a promising way of assessing the role of mergers in  black hole growth and evolution. 

\section{Acknowledgments}

A. E. thanks Fundaci\'on Andes and the Yale University Science Development Fund for fellowship support. D. M. gratefully acknowledge support from the Chilean Centro de Astrof\'isica  FONDAP 15010003.

\appendix



\end{document}